# Strain Modulated Electronic Properties of Ge Nanowires – A First Principles Study

Paul Logan,[1] Xihong Peng,[2,*]

[1]Department of Physics, Arizona State University, Tempe, AZ 85287

[2]Department of Applied Sciences and Mathematics, Arizona State University, Mesa, AZ 85212

* To whom correspondence should be addressed.  E-mail: xihong.peng@asu.edu.

## ABSTRACT

We used density-functional theory based first principles simulations to study the effects of uniaxial strain and quantum confinement on the electronic properties of germanium nanowires along the [110] direction, such as the energy gap and the effective masses of the electron and hole. The diameters of the nanowires being studied are up to 50 Å. As shown in our calculations, the Ge [110] nanowires possess a direct band gap, in contrast to the nature of an indirect band gap in bulk. We discovered that the band gap and the effective masses of charge carries can be modulated by applying uniaxial strain to the nanowires. These strain modulations are size-dependent. For a smaller wire (~ 12 Å), the band gap is almost a linear function of strain; compressive strain increases the gap while tensile strain reduces the gap. For a larger wire (20 Å - 50 Å), the variation of the band gap with respect to strain shows nearly parabolic behavior: compressive strain beyond -1% also reduces the gap. In addition, our studies showed that strain affects effective masses of the electron and hole very differently. The effective mass of the hole increases with a tensile strain while the effective mass of the electron increases with a compressive strain. Our results suggested both strain and size can be used to tune the band structures of nanowires, which may help in design of future nano-electronic devices.  We also discussed our results by applying the tight-binding model.

## 1. Introduction

One-dimensional semiconductor nanostructures, such as Si and Ge nanowires, have attracted extensive research efforts over the past decade.[1-12] They are expected to play important roles as both interconnects and functional components in future nanoscale electronic and optical devices, such as light-emitting diodes (LEDs),[3] ballistic field-effect transistors (FETs),[4,5] inverters,[6] and nanoscale sensors.[7,8] Therefore, it is of great importance to study electronic properties of those nanowires, such as band gap, density of state and effective mass of charge carriers. Compared to the material of Si, Ge has some superior properties. For example, Ge has an indirect band gap of 0.66 eV while the indirect band gap of Si is at a value of 1.12 eV. Ge also has higher electron/hole mobility, i.e. $\mu_n$ = 3800 cm$^2$V$^{-1}$s$^{-1}$ and $\mu_p$ = 1800 cm$^2$V$^{-1}$s$^{-1}$, compared to $\mu_n$ = 1500 cm$^2$V$^{-1}$s$^{-1}$ and $\mu_p$ = 450 cm$^2$V$^{-1}$s$^{-1}$ in Si at room temperature.[13-15] A much lower intrinsic resistivity of 46 Ω·cm in Ge is compared to 3.2×10$^5$ Ω·cm in Si. Therefore, Ge offers appealing opportunities for advanced device scaling, such as low drive voltage and high drive currents for high-speed electronics.[14] From the point of view of the nanoscale applications, the quantum confinement effects on Ge nanostructures are more prominent than on Si nanowires, which is essentially related to a much larger excitonic Bohr radius of 24.3 nm in Ge compared to 4.9 nm in Si. This makes Ge nanostructures with novel electronic properties more ready to be fabricated.

Recently, researchers were able to grow single crystals of Ge nanowires with diameters down to a few angstroms and lengths of tens of micrometers.[10, 15-18] In these nanoscale wires, the charge carriers, electrons or holes, are confined in the lateral direction of the wires, thus quantum confinement effect is expected to play an important role. This effect has been observed, for example, in photoluminescence (PL) studies, and found to exhibit substantial blue-shift of emission with reduction of nanowire diameter. For instance, Audoit et al.,[11] have grown Ge nanowires in the size range of 22 Å - 85 Å using supercritical fluid methods. They observed a clear blue-shift in the PL of the nanowires compared to the Ge band gap of 0.66 eV. Theoretically, researchers found that the band gap of Ge nanowires are dependent on several factors, such as size,[19-21] crystalline orientation,[20-22] surface chemistry,[22, 23] and doping.[22, 24]

However, there is only very limited study[11] of strain effects on electronic properties of Ge nanowires. It is well known that strain is a very important factor from the growth and application aspect of nano-devices. First, strain is not avoidable during epitaxial growth if there is a lattice mismatch between grown nanostructure and substrate. Second, in many applications such as nano-sensors, nanowires are usually embedded in some materials, within the coatings bring strains to the wires. In the field of microelectronics, strain has become a routine factor to engineer band gaps of semiconductors. Recently, people found that strain can enhance the device's performance by increasing the effective mass of the electron and hole.[25-27] Here, we give a thorough study of the strain effects on band structure of Ge nanowires with different sizes, using first principles calculations.

## 2. Simulation details

Our first principles density-functional theory (DFT)[28] calculations were performed using the Vienna computational code VASP.[29] The DFT local density approximation (LDA) was applied. In detail, we used a pseudo-potential plane wave approach with kinetic energy cutoff of 300.0 eV. The core electrons are described using ultra-soft Vanderbilt pseudo-potentials.[30] The dangling bonds in the Ge wire surface are saturated by hydrogen atoms. The size of the simulation cell along the axial direction of the [110] Ge nanowires is initially set to $a_{initial} = 3.977$ Å, taken from the lattice constant of bulk Ge 5.6245 Å (i.e. $a_{initial} = a_{bulk}/\sqrt{2}$). The lateral size of the cell is chosen so that the distance between the wire and its replica (due to periodic boundary conditions) is more than 8 Å. Under this configuration, the interactions between the wire and its replica are negligible. The [110] axial lattice constant is then optimized through the technique of total energy minimization. The electronic properties of the wire, such as the band gap, the effective mass of charge carrier, are then calculated by solving the Kohn-Sham equation within the frame of DFT. The band gap of a wire is defined by the energy difference between the bottom of the conduction band (conduction band edge – CBE or the lowest unoccupied molecular orbital – LUMO) and the top of the valence band (valance band edge – VBE or the highest occupied molecular orbital – HOMO). The effective masses of the electron and hole can be readily calculated according to the formula $m^* = \hbar^2 (d^2E/dk^2)^{-1}$ from the band structure of the wire.

Table 1 lists the Ge wires studied in the present work. $N_{Ge}$ is the number of Ge atoms in a given wire; $N_H$ represents the number of H atoms needed to saturate the surface dangling bonds in the wire; D is the diameter of a wire in the unit of Å, defined as the longest distance between two Ge atoms in the cross-section of the wire. Fig. 1 gives the snapshots of two Ge nanowires at size of 18 Å and 30 Å viewed from the cross-section and the side of the wire. Blue dots are Ge atoms and white are H atoms.

Based on the relaxed configurations of a wire (with the axial lattice constants listed in Table 1), we then applied uniaxial tensile/compressive strain by scaling the axial lattice constant of the wire. The positive values of strain refer to uniaxial expansion, while negative corresponds to compression (note that the lateral x- and y-coordinates of the wire are further optimized at a given strain). Our study showed electronic properties of the wire are affected significantly by strain.

## 3. Results and data analyses

### I. Axial lattice constant

We first characterized the geometries of the relaxed Ge nanowires. The lattice constant $a_{bulk}$ in bulk Ge is 5.6245 Å based on the simulation parameters mentioned above. The initial axial lattice constant $a_{initial}$ = 3.977 Å of a [110] wire is obtained from bulk Ge. This axial lattice constant was defined as the interplanar distance between two consecutive [110] planes. The total energy of the wire was calculated by relaxing all lateral $x$- and $y$-coordinates. In order to optimize the axial lattice constant (along $z$-direction), we performed a series of calculations of the total energy with different lattice constants for a given wire. Then the total energy in the wire was plotted as a function of the axial lattice constant. Through a parabolic fitting of the above plot (the total energy versus lattice constant), we were able to find the optimized axial lattice constant $a_{optimized}$. For example, we found that $a_{optimized}$ = 3.997 Å for the wire with a diameter of 12 Å, which is greater than $a_{initial}$ = 3.977 Å. That means the wire expands along the axial direction upon relaxation, which is in a good agreement with experimental data[11]. For all wires studied in present work, $a_{optimized}$ are reported in the fifth column of Table 1. For instance, $a_{optimized}$ = 3.984 Å for the second smallest wire of 18 Å. Larger wires beyond 20 Å have the same lattice constant of $a_{optimized}$ = 3.977 Å, as well as the bulk. This implies the axial expansion is negligible in the wires with a diameter larger than 20 Å.

In our previous work of Si [110] nanowires,[31] we also found the Si nanowires expanded axially upon relaxation. However, the expansion only became negligible when the size of the nanowires was beyond 40 Å. Although both Si and Ge crystals have diamond structures and tetrahedral networking, a Si-Si bond is stronger than a Ge-Ge bond. When the bonds of surface atoms of Si nanowires are cut off, it will cost more surface energy compared with the case for the Ge nanowires. This may account for the larger required size of the Si nanowires for the disappearance of axial expansion, in which extra surface energy is accommodated by the interior and saturation atoms without much change in the lattice constant.

### II. Band gaps

#### a) Size effects

Bulk Ge is an indirect band gap material with the conduction band minima located at L along the [111] direction. However, if the Ge nanowires are along the [110] direction, they will show a direct band gap at Γ, as shown in the literature.[19, 21, 22] In Fig. 2, we present the band structures of Ge nanowires with varied diameters. It clearly demonstrates a direct band gap – both CBE and VBE located at Γ, consistent with previous work. [19, 21, 22]

As mentioned before, the band gap of a Ge wire is defined by the energy difference between CBE and VBE (or HOMO-LUMO gap). In the fifth column of Table 1, we report the DFT predicted band gaps for the Ge wires. It is known that DFT underestimates band gaps of semiconductors, while advanced GW method[32-34] and quantum Monte Carlo calculations[35-37] provide improved predictions. However, previous studies[38] on Si nanoclusters and nanowires showed that the DFT gap predicts a similar size-dependency as the optical gap obtained using GW and quantum Monte Carlo methods. [35, 37] The band gap of the Ge nanowire in Table 1 is increased when the size of the wire is reduced. This effect is primarily due to quantum confinement. Our predicted size-dependence of the band gap in Ge nanowires is in a good agreement with the literature[19] and reference within.

### b) Strain effects

It is also interesting to observe that the band structures are modulated with strain. For example, in Fig. 3-(a), we compared the band structures of the Ge nanowire with a diameter of 18 Å, with and without strain. Black solid lines are the band structure without strain; red dashed lines are under tensile uniaxial strain; blue dotted lines are under compressive uniaxial strain. Generally, strain has dominant effects on the band structure near Γ (i.e. energy is shifted evidently with strain, see the dashed pink ovals), while it has negligible effects on wave vectors far away from Γ (i.e. minimal energy shift under strain, see the solid green ovals). Most electronic properties are related to the bottom of the conduction band and the top of the valence band. Therefore, the energy variation of these two edges was particularly singled out and presented in Fig. 3-(b) and -(c). From those two figures we can clearly see that strain modifies the energies of CBE and VBE dramatically near Γ, and has negligible energy shifts on wave vectors far away from Γ.

The variation of band gaps as a function of uniaxial strain for several different sized wires is plotted in Fig. 4. For the wire with a diameter of 12 Å, the band gap variation with strain is almost linear, as shown by the green-triangle curve. The gap decreases with expansion and increases with compression. The gap variation with strain in the wire with diameter of 18 Å, shown by the blue-star curve, has a more modest change in the gap for a given strain, compared with the 12 Å wire. However, for the wire with diameter of 25 Å, the gap variation with strain, shown by the red-diamond curve, exhibits a nearly parabolic behavior, the gap drops not only under expansion, but also under compression beyond 2%. This parabolic behavior is more evident for the larger wire with diameter of 37 Å, shown by the black-dot curve. We conclude that the strain effect on the band gap in Ge wires is strongly dependent on its size.

## III. Effective masses

### a) Size effects

The effective masses of the electron and hole can be calculated from the band structure of the Ge wires. We take the nanowire with a diameter of 37 Å as an example. As shown in Fig. 5, we first calculated the energy dispersion curve without strain near Γ in a fine step, shown by the solid and hollow circles. The wave vector ranges from -0.1 to +0.1, where ± 0.1 is in units of $2\pi/a$ ($a$ is the axial lattice constant). Figure 5 also shows the energy dispersion curves under different strains, which will be discussed later. Then the curves of the energy-dispersion around Γ are fitted using the second order polynomial $E = C_1 k^2 + C_2 k + C_3$. We can obtain the curvature of the energy-dispersion-curve as $C_1 = \frac{1}{2}(d^2 E / dk^2)$. Furthermore, we can calculate effective mass of the electron and hole through the relation $m^* = \hbar^2 / 2C_1$. In Table 1, we report the calculated results in the last two columns. $m_e^*$ represents the effective mass of the electron, while $m_h^*$ is the effective mass of the hole, in units of electron mass $m_e$. For example, the effective mass of the electron $m_e^*$ in the wires with diameters of 12 Å and 18 Å are $0.12\,m_e$; in the larger four wires are $0.11\,m_e$, with negligible change. On the other hand, the effective mass of the hole $m_h^*$, in general, increases with size, from $0.11\,m_e$ in the wire with diameter of 12 Å to $0.47\,m_e$ in the 47 Å wire. Note that the smaller effective mass of the charge carrier in a material implies larger mobility of charge carrier, thus increasing the operating speed of devices made from the material.

### b) Strain effects

We studied the effect of strain on effective masses of the electron and hole near Γ. In order to calculate the effective masses of the wire with diameter of 37 Å, the dispersion relation in the region near Γ are plotted under different values of strain, as shown in Fig. 5. As shown in Fig. 6-(a) and -(b) by the green-star curves, under 2% compressive strain, the effective mass of the electron is increased to $0.166\,m_e$ (increased by 55%), while the effective mass of the hole is reduced to $0.133\,m_e$ (decreased by 57%). In contrast, under 2% expansive strain, the effective mass of the electron is deceased to $0.102\,m_e$ (reduced by 4.7%), while the effective mass of the hole increases dramatically to $1.139\,m_e$ (increased by 270%), resulting from the nearly flat energy dispersion relation in Fig. 5, shown by the blue-pentagon curve.

The change of effective masses of the electron and hole with strain is also dependent on the size of nanowires, given in Fig. 6-(a) and -(b). However, all these changes have a general trend. It shows, in Fig. 6-(a), that the effective mass of the hole reduces under compression, while enhanced dramatically with tensile strain. However, the effective mass of the electron increases rapidly with compressive uniaxial strain, while decreasing mildly with tensile strain, as shown in Fig.6-(b).

4. **Discussion**

   A) **Size-dependence of strain effects on the gap**

In order to understand the size-dependence of the strain effects on the gap (Fig. 4), we first examined the variations of the energies of VBE and CBE with strain. The energies of CBE and VBE in two wires, whose diameters are 12 Å and 37 Å, are plotted as a function of strain in Fig. 7-(a) and -(b). It is clear that the energies of the CBE and VBE in the 12 Å wire are a linear function of strain. The energies of the CBE and VBE decrease with expansion while increasing with compression. In addition, the slope of the CBE plot, shown by the hollow-star graph, is slightly smaller (i.e. more negative) than that of the VBE plot, given by the graph of solid stars. Since the band gap is given by the energy difference between the CBE and VBE, it is also a nearly linear function of strain (see 12 Å curve in Fig. 4). However, for the 37 Å wire, the energies of the CBE and VBE, shown by the graphs of the hollow and solid dots, are not linear functions with strain. Generally, both energies of the CBE and VBE are reduced under expansion and increased with compression. However, the curve of the CBE decreases faster than that of the VBE under expansion. On the other side, the curve of the CBE increases slower than that of the VBE under compression.

To understand the behaviors of the strain effects on CBE and VBE shown in Fig.7, it is necessary to study the strain response in the lateral directions (*i.e. x*- and *y*-directions) in the wire when strain is applied to the axial direction (*i.e. z*-direction). As it would be expected, once the axial strain is applied, the bonds in the *x*- and *y*-directions will change due to the Poisson effect. For example, for the 12 Å wire, the *x*- and *y*-directions shrink 0.1% and 0.5%, respectively, when 1.5% expansion is applied to the *z*-direction. Similarly, for the 37 Å wire, the *x*- and *y*-directions reduce by 0.2% and 0.4%, respectively, when 2% expansion is applied to the *z*-direction.

The electron wavefunction contour plots at the iso-value of 0.02 for the VBE and CBE from the views of the lateral cross-section and the side in the 12 Å Ge wire are presented in Fig. 8. For both views of the wire, the orbitals of the VBE and CBE have bonding character – the electron cloud is mainly located in the intermediate regions shared by Ge atoms. From the above discussion of strain response, the lateral *xy*-plane will bear compressive strain once expansive axial strain is applied to the wire. That means in the *xy*-plane the distance of Ge atoms will be reduced. The reduction of Ge-Ge bond lengths makes the electron cloud of the VBE and CBE orbitals more efficiently shared by Ge atoms. This effect results in an increased electron-nucleus Coulomb attraction, thus an appreciable decrease of energies of both the VBE and CBE (the change in the electron-electron repulsion energy is relatively small). In contrast, with uniaxial compression, the lateral *xy*-plane experiences expansive strain. With this expansion, energies of both the VBE and CBE increase due to the decrease of electron-nucleus attraction. This explains the general variation trends of the energies of the VBE and CBE with respect to strain in Fig. 7 – *i.e.* the energies of the VBE and CBE increase with compression while decreasing with expansion. In addition, from Fig. 8, we found that the orbital of the CBE is more delocalized than that of the VBE. Thus, the electron cloud of the CBE is more effectively shared by Ge atoms in the *xy*-plane compared to that of the VBE. As a result, the energy of the CBE is more sensitive to strain than that of the VBE. Therefore, the slope of the CBE curve in the 12 Å wire in Fig. 7 is slightly larger than that of the VBE curve.

For the 37 Å wire in Fig. 7, we found the curve of the CBE decreases faster than that of the VBE under expansion, while the curve of the CBE increases slower than that of the VBE under compression. This can be understood from the combined effects of strain and degeneracy of band edges. If we only consider the effect of strain in the larger nanowire, we will expect a similar linear variation of band edges with strain as discussed for the small 12 Å wire. However, for the larger wire, the band edges are degenerate due to the tetrahedral (Td) symmetry of the core Ge atoms. Under uniaxial strain, the Td symmetry of the core Ge atoms is broken and the degeneracy of the band edges is released. In this case, the degeneracy lifting of band edges will make the energies of the CBE and VBE vary as parabolic functions of strain [39]. In this parabolic behavior, the energy of the CBE decreases while that of the VBE increases under both expansion and compression (see Reference 39). Thus the curves in Fig. 7 for the larger wire can be understood from the combined effects of strain and degeneracy lifting of band edges.

**B) Strain and size effects on the effective masses**

As shown in Fig. 3-(b) and -(c), the strain effect on the electronic bands is prominent at the gamma point while this effect becomes much smaller as approaching the K-edge of the Brillouin zone, i.e. the X point. In order to understand this result, we applied the tight-binding model. [40] In this model, the wave function of a crystal is in a form of a Block function, and the energy of the band can be expressed as $E(k) = E_v - \beta - \gamma \sum_{n,n} \cos(\vec{k} \cdot \vec{R})$, where the summation goes over those $\vec{R}$ of the nearest neighbors. To discuss the strain effect on the Ge nanowires, the energy can be further simplified as $E(k) = E_v - \beta - 2\gamma \cos(\frac{k_{//} a}{2})$, where $E_v$ is the energy of atomic orbitals, $k_{//}$ is the magnitude of wave vector along the direction of the wire axis, $\beta$ is a small quantity contributed by the energy correction near the nucleus position, $\gamma$ is called the overlap integral and is another term of energy correction dependent on the overlap between orbitals centered at two neighboring atoms, and *a* is the lattice constant. For the gamma point ($k_{//} = 0$), the energy is $E(\Gamma) = E_v - \beta - 2\gamma$. For the X point ($k_{//} = \frac{1}{2}\frac{2\pi}{a}$), the contribution from the overlap integral γ vanishes and the energy is $E(X) = E_v - \beta$. By applying strain to a Ge nanowire, the bond length between Ge atoms will be changed. Thus we expect a prominent modification of the γ value, while the variation of β is negligible due to its local nature. Referring to the above formulae of $E(\Gamma)$ and $E(X)$, strain will bring a more pronounced effect in the energy at Γ, compared to other K points.

It is also interesting to notice the size-dependence of the effective masses of the electron and the hole as shown in Table 1. We found that the effective mass of the hole decreases substantially with the reduced diameters of the Ge nanowires. In contrast, the effective mass of the electron is less sensitive to the size. Karanth and Fu [41] showed the similar findings in their calculations of InP nanowires. In order to understand the simulated results we need consider the quantum confinement effect on the nanowires. As the diameter of Ge nanowire is reduced, the component of wave vector perpendicular to the wire axis, $k_\perp$, becomes quantized and inversely proportional to the size of the nanowire. [42] In this $k_\perp$ always has a finite value. As a result, the top/bottom of the valence/conduction bands will shift away from its bulk position. This causes a non-parabolic band curvature, enhancing the effective mass of electron. [42] This effect of non-parabolicity is originated from the second order perturbation and is usually small, consistent with our calculations of the effective mass of the electrons in Table 1. For the hole, the situation becomes different. In Ge bulk crystal, the valence band is degenerated with the light hole and heavy hole bands at Γ. When $k_\perp$ becomes quantized, this degeneration will be released. The energy of the heavy hole band may shift lower, compared with the energy of the light hole band. The reason could be that for the heavy hole band, the overlap of wave function in the direction perpendicular to the wire axis is significant. [43] In contrast, for the light hole band, the overlap in the perpendicular direction is small (see Fig. 8 (a)) although this overlap is significant along the quantum wire axis as shown in Fig. 8 (b). A larger overlap of wave function in the perpendicular direction implies a smaller effective mass in this direction ($m_{\perp eff}$). Since here the amount of energy down-shift by the quantized $k_\perp$ is approximately $\frac{\hbar^2 k_\perp^2}{m_{\perp eff}}$, [43] the energy of the heavy hole band, which has a smaller $m_{\perp eff}$, will be decreased more. [43]. Consequently, the light hole band becomes the very top valence band at Γ of the nanowire and gives a smaller effective mass of the hole, compared with the value of bulk crystal.

Finally it is necessary to briefly discuss the impact of surface passivation on the results. In the present work, the surface dangling bonds of Ge wires are passivated by hydrogen atoms. From the contour plots of electronic wave functions near the Fermi level, the orbitals including HOMO and LUMO are mainly contributed by Ge atoms rather than H [44], see Fig. 8-(a) through (d). We conclude that our results of band gap and effective masses of Ge nanowires are predominantly dependent on the diameter and strain, rather than the surface H atoms. Experimentally, the surface of Ge nanowires may be saturated by oxygen under an ambient condition. From previous studies of Si nanowires and quantum dots, [35, 45-47] this oxygen shell would bring surface states near the Fermi level to become the new HOMO and LUMO of the system. All electronic properties related to the new HOMO and LUMO are expected to be different from that of H passivation. This effect of surface chemistry (beyond the scope of the present work) could be another tuning factor to modulate the electronic properties of semiconductor nanostructures.

## 5. Conclusion

In summary, we found that (1) the nanowires expand along the axial [110] direction compared to bulk Ge: the expansion is evident for small wires with diameter less than 20 Å; (2) the band structures of Ge [110] wires display a direct band gap at $\Gamma$; (3) the band gap variation with uniaxial strain is size-dependent: for smaller wires with size around 12 Å, the band gap is a linear function of strain while for the wires in the range of 20 Å - 40 Å, the gap variation with strain shows a nearly parabolic behavior resulting from the localized nature of band edges; (4) strain affects the effective masses of the electron and hole in a different manner: expansion increases the effective mass of the hole, while compression increases the effective mass of the electron; (5) the strain and size effects on these electronic properties of nanowires may be understood by applying the tight-binding model. Our studies show that the effective masses of the electron and hole can be reduced by tuning the diameter of the wire and applying appropriate strain, which supports the motivation for using Ge nanowires as components and interconnects in future nanoelectronics.

**Acknowledgement**

This work is supported by the Research Initiative Fund from Arizona State University (ASU) to Peng. We are very thankful to the following for providing computational resources: ASU Fulton High Performance Computing Initiative (Cluster Saguaro), National Center for Supercomputing Applications and Pittsburgh Supercomputing Center. Jeff Drucker and Fu Tang are acknowledged for helpful discussions. Fu Tang, Heather Canary, Sonia Vega Lopez, Claire Lauer, Gayle Stever, Kathleen Woolf, and Keith Martin are acknowledged for review of the manuscript.

**Table captions**

Table 1, A list of studied Ge nanowires along [110] direction in present work. $N_{Ge}$ is the number of Ge atoms in a given wire; $N_H$ represents the number of H atoms needed to saturate the surface dangling bonds; D is the diameter of a wire; the fourth column is the optimized axial lattice constants of the wires; $E_g$ is DFT predicted band gaps; $m_e^*$ and $m_h^*$ are DFT predicted effective masses of the electron and hole.

**Figure captions**

Fig. 1  (Color online) Snapshots of Ge nanowires with size of 18 Å (top) and 30 Å (bottom) viewed from the wire cross-section (left) and the side (8-contiguous simulation cells along the axial *z*-direction). Blue dots are Ge atoms, white are H atoms.

Fig. 2  The band structures of Ge nanowires with varied diameter along [110] direction. They show a direct band gap located at Γ. Fermi level is set to zero in all cases.

Fig. 3 (Color online) (a) The band structures of Ge [110] nanowires with a diameter of 18 Å, with and without strain.  Black solid lines are the band structure without strain; red dashed lines are under tensile uniaxial strain; blue dotted lines are under compressive uniaxial strain. The energy variations of the bottom of the conduction band (b) and the top of the valence band (c) in Ge nanowires of 18 Å with uniaxial strain. The uniaxial strain has a dominant effect of shifting energies on the conduction and valence bands near Γ.

Fig. 4  (Color online) The change of the DFT predicted band gap in Ge wires as a function of uniaxial strain ε at different size. Positive strain refers to uniaxial expansion while negative strain corresponds to its compression.

Fig. 5  (Color online) The conduction and the valence bands of wire with diameter of 37 Å  at the near region of Γ are plotted under different values of uniaxial strain. The effective masses of the electron and hole are obtained through parabolic fitting the band edges according to the formula $m^* = \hbar^2 (d^2 E / dk^2)^{-1}$.

**Fig. 6** (Color online) The change of effective masses of the electron (left) and hole (right) are plotted as a function of uniaxial strain for nanowires at different size. It shows that the effective mass of the electron increases rapidly with compressive uniaxial strain, while decreasing mildly with tensile strain. However, the effective mass of the hole reduces under compression, while enhanced dramatically with tensile strain.

**Fig. 7** (Color online) The changes of CBE and VBE energies in Ge nanowires are plotted as a function of uniaxial strain.

**Fig. 8** (Color online) Electron wavefunction contour plots at the iso-value of 0.02 for the VBE (left) and CBE (right) in the 12 Å Ge nanowire viewed from the lateral *xy*-plane (top) and the side *yz*-plane (bottom). Red and green colors correspond to positive and negative values of the wavefunctions. Blue dots are Si atoms, white are H atoms.

| $N_{Ge}$ | $N_H$ | D (Å) | Axial Lattice (Å) | $E_g$ (eV) | $m_e^*$ | $m_h^*$ |
|---|---|---|---|---|---|---|
| 16 | 12 | 12 | 3.997 | 1.54 | 0.12 | 0.11 |
| 42 | 20 | 18 | 3.984 | 1.02 | 0.12 | 0.09 |
| 76 | 28 | 25 | 3.977 | 0.73 | 0.11 | 0.15 |
| 110 | 32 | 30 | 3.977 | 0.61 | 0.11 | 0.19 |
| 172 | 44 | 37 | 3.977 | 0.49 | 0.11 | 0.31 |
| 276 | 52 | 47 | 3.977 | 0.39 | 0.11 | 0.47 |

**Table 1, Peng *et al*.**

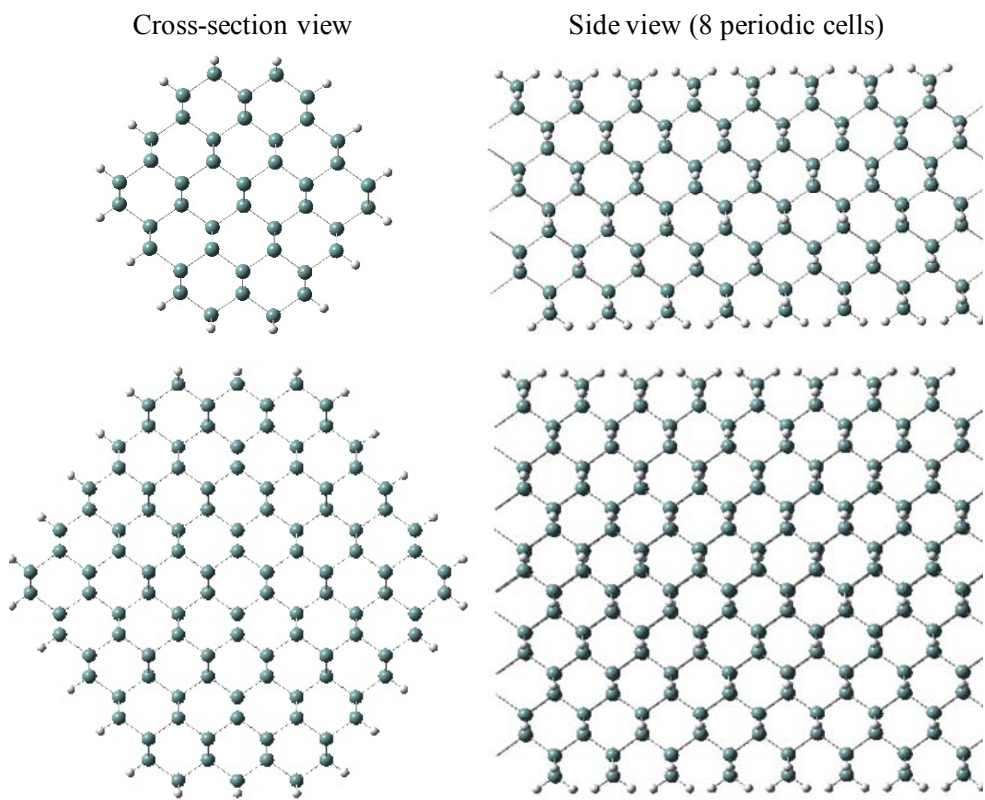

**Fig. 1, Peng *et al*.**

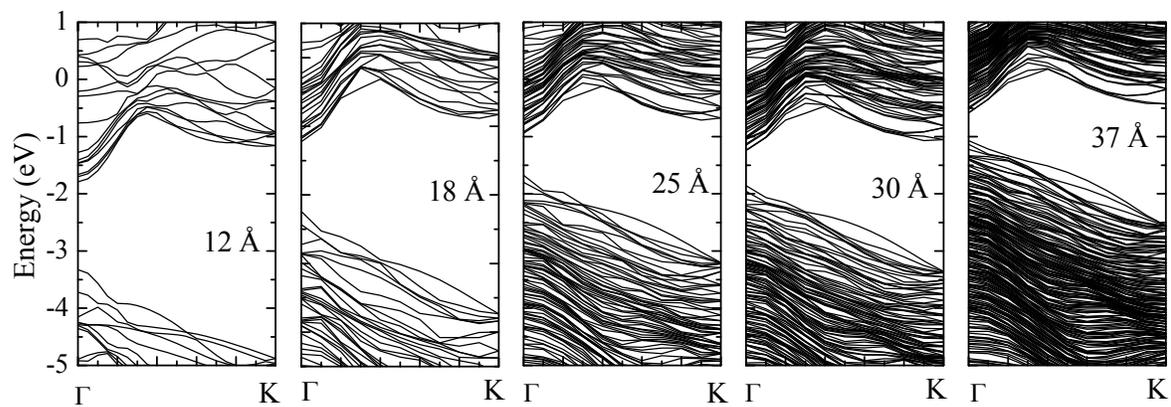

**Fig. 2, Peng *et al*.**

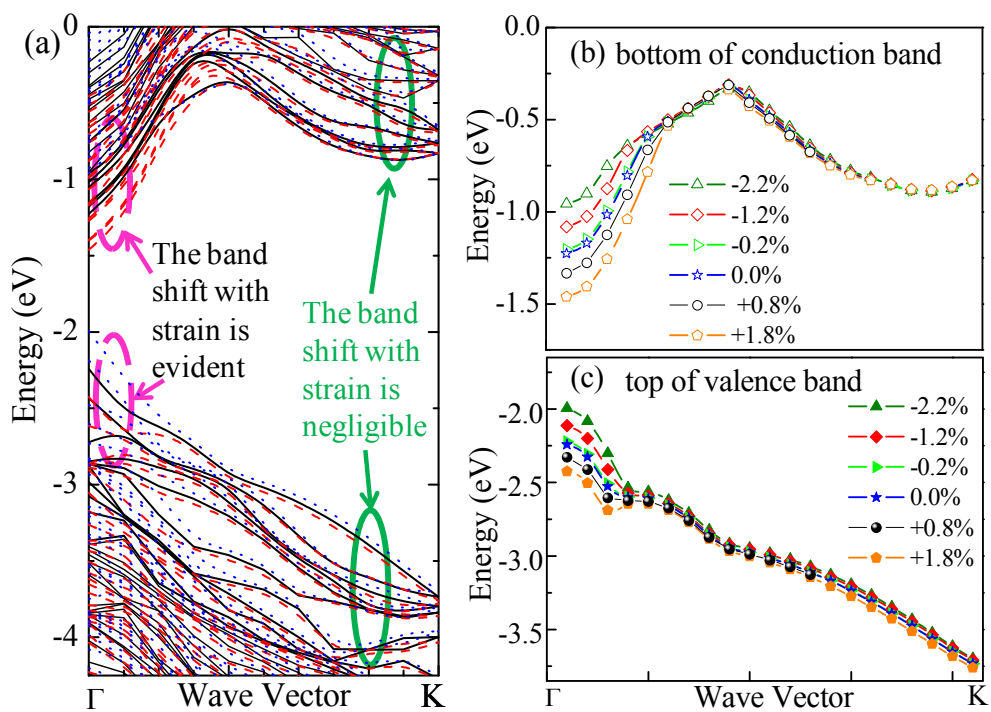

**Fig. 3,** Peng *et al*.

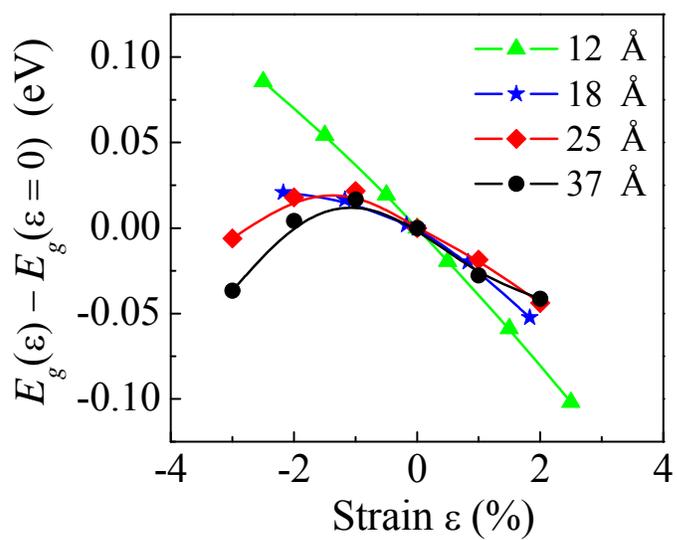

**Fig. 4,** Peng *et al*.

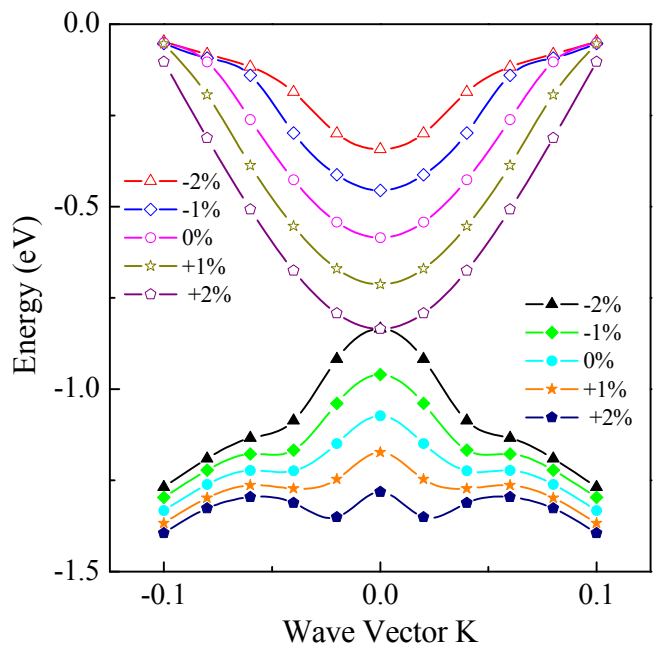

**Fig. 5,** Peng *et al*.

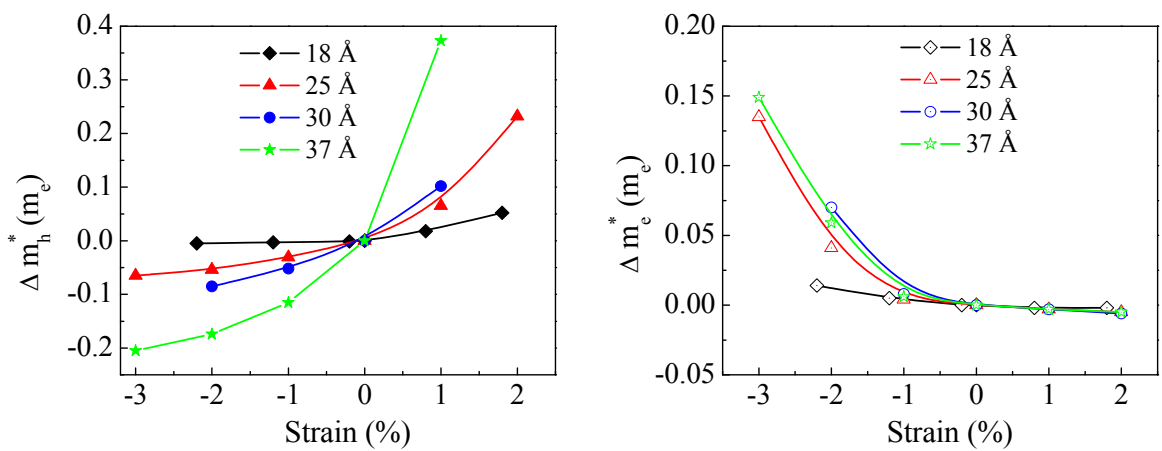

**Fig. 6, Peng *et al*.**

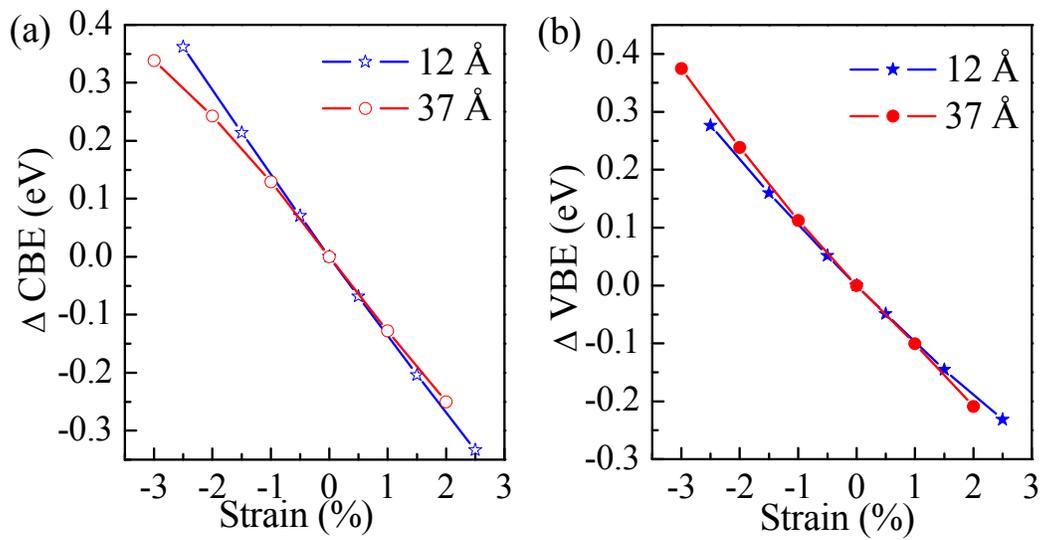

**Fig. 7,** Peng *et al*.

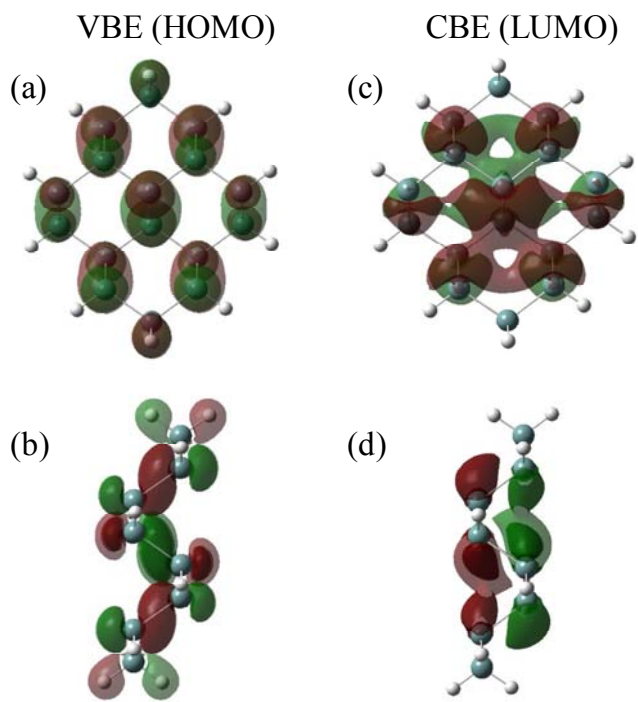

**Fig. 8,** Peng *et al*.